# Moist convection scheme in Model E2

Daehyun Kim[1], Anthony D. Del Genio[2], and Mao-Sung Yao[2,3]

[1]*Lamont-Doherty Earth Observatory of Columbia University, Palisades, NY*
[2]*NASA Goddard Institute for Space Studies, New York, NY*
[3]*Sigma Space Partners, Institute for Space Studies, New York, NY*

This documentation describes the moist convection scheme implemented in the NASA Goddard Institute for Space Studies General Circulation Model - Model E2.

1) Convective cloud model

The moist convection scheme in Model E2 uses the entraining-detraining plume model for updrafts and downdrafts. For each updraft or downdraft plume, the following equations are used to diagnose the properties of the plume.

Mass flux ($M$):

$$\frac{\partial M^{u,d}}{\partial z} = (\varepsilon - \delta) M^{u,d}, \qquad (1)$$

where superscripts $u$ and $d$ indicates the updraft and downdraft, respectively. Here, $\varepsilon$ and $\delta$ represent fractional entrainment and detrainment rate, respectively.

Potential temperature ($\theta$), specific humidity ($q$), and horizontal momentum ($u, v$):

$$\frac{\partial M^{u,d}\emptyset^{u,d}}{\partial z} = \varepsilon M^{u,d}\overline{\emptyset} - \delta M^{u,d}\emptyset^{u,d} + S_{\emptyset^{u,d}}, \tag{2}$$

where the overbar indicates a gridbox-mean quantity. The source/sink term $S$ for each variable $\emptyset$ is summarized in Table 1. When condensation occurs, the resulting condensate is used in cumulus microphysics described in the next section.

Note that the grid-scale budget equation for any generic prognostic variable $\emptyset$ is given by,

$$\frac{\partial \emptyset}{\partial t} = -\frac{1}{\overline{\rho}}\frac{\partial}{\partial z}\left[M^u\emptyset^u + M^d\emptyset^d - (M^u + M^d)\overline{\emptyset}\right] + S_{\emptyset^u} + S_{\emptyset^d}. \tag{3}$$

For the updraft plume, the vertical velocity of the cloud parcel ($w$) is calculated following Gregory (2001):

$$\frac{1}{2}\frac{\partial (w^u)^2}{\partial z} = agB^u - b\delta(w^u)^2 - \varepsilon(w^u)^2, \tag{4}$$

where $a = 1/6, b = 2/3$, $g$ is gravitational acceleration and $B$ is parcel buoyancy. At each level, the buoyancy of the cloud parcel is diagnosed using

$$B^u = \frac{1}{T_v^a}(T_v^u - T_v^a) - \mu^u, \tag{5}$$

where $T_v$ is virtual temperature and $\mu$ is the adiabatic cloud water mixing ratio.

The fractional entrainment and detrainment rate used in (1), (2), and (4) are determined differently in the updraft and downdraft. In updraft calculations, if the buoyancy of the parcel is positive, the fractional entrainment rate is determined by,

$$\varepsilon = \frac{C_\varepsilon a g B^u}{(w^u)^2}, \tag{6}$$

where $C_\varepsilon$ is a constant introduced by Gregory (2001) as a fraction of the kinetic energy gained by buoyancy that is transferred to the air mass entrained from the environment. The Model E2 convection scheme launches two plumes with different values of $C_\varepsilon$ to allow for instantaneous variability of convection depth within a gridbox (e.g., deep and shallow, deep and congestus, etc.). Above the level of neutral buoyancy, the model no longer entrains, but it detrains convective air mass to the environment and assumes the fractional detrainment rate to also given by (6), but multiplied by −1. The fractional entrainment rate for the downdraft is set to $2\times10^{-4}$ (m⁻¹); there is no downdraft vertical velocity equation. The downdraft does not detrain until it becomes positively buoyant, with 75% of its mass detraining at each subsequent.

The updraft calculation is triggered from a conditionally unstable layer where the virtual moist static energy of the layer exceeds the saturation virtual moist static energy of the level above (Yao and Del Genio 1989). The mass flux at cloud base is determined as the mass flux required to remove the instability at cloud base during a convective adjustment time currently specified to be 1 hour. This total mass flux is partitioned into two parts for the less-entraining (i.e. smaller $C_\varepsilon$ value) and more-entraining plumes, depending on the grid-scale sigma velocity at the level above. The stronger the upward motion at the level above cloud base, which represents lower-level convergence, the larger the mass fraction of the less-entraining plume. For the potential temperature, specific humidity, and horizontal momentum of the updraft, gridbox-mean values at the cloud-base level are used. The updraft vertical velocity is initialized using the turbulent kinetic energy (TKE) from the planetary boundary layer scheme ($2\sqrt{\frac{2}{3}\text{TKE}}$ for less-entraining plume and $\sqrt{\frac{2}{3}\text{TKE}}$ for more-entraining plume).

A downdraft can be triggered, during the course of calculations of the rise of the updraft plume, from any level where an equal mixture of cloudy and environmental air has negative buoyancy. A downdraft calculation is initialized with a fixed fraction (1/6) of the mass of the updraft and an equal amount of mass from the environment. For potential temperature, specific humidity, and horizontal momentum, cloudy and environmental quantities are averaged for the initial downdraft properties. Evaporation of convective condensate produced and carried by the updraft modifies the potential temperature and specific humidity of the downdraft plume. In the AR5 version, all condensate can evaporate in the downdraft until it saturates. The downdraft is allowed to descend below cloud base if it is negatively buoyant. (In AR5 version the buoyancy is simply based on the temperature difference between the downdraft plume and environment; the buoyancy was changed to include effects of water vapor and convective condensate in later versions).

2) Cloud microphysics

At each level, the fractions of condensate to be i) advected upward, ii) detrained out of the updraft, and iii) precipitated are diagnosed based on the assumed cloud particle size distribution, terminal velocity of particles, and vertical velocity of the updraft from Eq. (4), following Del Genio et al. (2005). Note that the updrafts are currently the only transport mechanism for condensate. Specifically, mass distribution ($M(D)$) of the condensate is assumed to be given by the Marshall-Palmer size distribution,

$$M(D) = (\pi \rho_w N_0/6) D^3 e^{-\lambda D}, \qquad (A6)$$

where $D$ is the particle diameter, $N_0 = 8 \times 10^6 m^{-4}$, $\lambda = (\pi \rho_w N_0/\mu)^{1/4}$, $\rho_w = \rho_l, \rho_g, \rho_i$ represents the densities of liquid ($\rho_l$), graupel ($\rho_g$) and ice ($\rho_i$).

Based on parcel temperature, different formulas are used for liquid, graupel, and ice/snow to calculate the amount of condensate that precipitates, detrains, and is advected upward

$$f_i = 1 - \frac{T_u - T_{ig}}{T_f - T_{ig}}, \tag{A7}$$

where $f_i$ is the fraction of frozen condensate existing as ice/snow (rather than graupel), $T_u$ is parcel temperature, $T_f$ is temperature at the freezing level, $T_{ig} = T_f - 4w_{frz}$ is the temperature where all condensate becomes ice/snow, and $w_{frz}$ is the vertical velocity of the parcel at the freezing level. Above 0° all convective condensate is assumed to be liquid.

Particle size-fall speed relationships are fits to the terminal velocity ($v_t$) measurements from field experiments, adjusted for pressure variations with respect to surface pressure $P_0$, given by

$$v_{tl}(D) = (-0.267 + 5.15 \times 10^3 D - 1.0225 \times 10^6 D^2 + 7.55 \times 10^7 D^3)\left(\frac{p_0}{p}\right)^{0.4}, \tag{8}$$

$$v_{tg}(D) = 19.3 D^{0.37} \left(\frac{p_0}{p}\right)^{0.4}, \tag{9}$$

$$v_{ti}(D) = 11.72 D^{0.41} \left(\frac{p_0}{p}\right)^{0.4}. \tag{10}$$

for liquid, graupel, and ice/snow, respectively.

Equations above are solved for the critical values of diameter $D_{c\pm}$ at which $v_{t\pm} = w^u \pm \Delta w$, where $\Delta w = \frac{1}{2}\frac{\Delta z}{\Delta t}$, $\Delta z$ is layer thickness, and $\Delta t$ is the model physics time step. A particle whose Doppler vertical velocity $v_t - w^u$ would not carry it out of the layer in one physics time is detrained, while particles with positive (negative) Doppler vertical velocity are advected upward (precipitated).

The amount of condensate in each category is calculated by integrating the Eq. (6) over the appropriate particle size range. For example, the mass of precipitating condensate (the part of the mass distribution with $D > D_{c+}$), which is

$$\mu_p = \left(\frac{\pi \rho_w N_0}{6}\right) \int_{D_{c+}}^{\infty} D^3 e^{-\lambda D} dD = (\pi \rho_w N_0/6) e^{-\lambda D_{c+}} \lambda^{-4} \left(D_{c+}^3 \lambda^3 + 3 D_{c+}^2 \lambda^2 + 6 D_{c+} \lambda + 6\right). \tag{12}$$

Any precipitating condensate that is not evaporated in the downdraft process can be re-evaporated in the environment if possible.

Table 1. Source/sink terms in Eq. (2).

| ∅ | θ | q | u, v |
|---|---|---|---|
| Source/sink | Latent heat of condensation/evaporation, freezing/melting | Evaporation/condensation | Convective-scale pressure gradient force |